\documentclass[9pt,twocolumn,twoside]{pnas-new}

\templatetype{pnasresearcharticle} 

\title{Experimental observation of the marginal glass phase in a colloidal glass}

\author[a,1]{Andrew P. Hammond}
\author[a]{Eric I. Corwin}

\affil[a]{Materials Science Institute and Department of Physics, University of Oregon, Eugene, Oregon 97403}

\leadauthor{Hammond} 

\significancestatement{

Glass is one of the easiest materials to work with, capable of forming complex and sophisticated structures with relative ease in the hands of a glass blower.  However, the underlying physical principles that allow glass to slowly melt and flow are among the least understood.  Recent theoretical breakthroughs hope to reveal a new kind of underlying simplicity within glasses, one which might extend to all manner of disordered and complex materials.  By studying the formation of a model glass system at extremely short time scales and small length scales we have proven the central prediction of these theories, that glasses undergo an extremely subtle transition while forming which controls their physical properties. 

}

\authorcontributions{A.H. and E.C. designed research, contributed analytic tools, analyzed data and wrote the paper; A.H. performed research.}
\authordeclaration{The authors declare no conflict of interest}

\correspondingauthor{\textsuperscript{1}To whom correspondence should be addressed. E-mail: ahammon7@uoregon.edu}

\begin{abstract}
The replica theory of glasses predicts that in the infinite dimensional mean field limit there exist two distinct glassy phases of matter: stable glass and marginal glass.  We have developed a technique to experimentally probe these phases of matter using a colloidal glass.  We avoid the difficulties inherent in measuring the long time behavior of glasses by instead focusing on the very short time dynamics of the ballistic to caged transition.  We track a single tracer particle within a slowly densifying glass and measure the resulting mean squared displacement (MSD).  By analyzing the MSD we find that upon densification our colloidal system moves through several states of matter.  At lowest densities it is a sub-diffusive liquid.  Next it behaves as a stable glass, marked by the appearance of a plateau in the MSD whose magnitude shrinks with increasing density.   However, this shrinking plateau does not shrink to zero, instead at higher densities the system behaves as a marginal glass, marked by logarithmic growth in the MSD towards that previous plateau value.  Finally, at the highest experimental densities the system returns to the stable glass phase.  This provides direct experimental evidence for the existence of a marginal glass in 3d.
\end{abstract}

\dates{This manuscript was compiled on \today}
\doi{\url{www.pnas.org/cgi/doi/10.1073/pnas.XXXXXXXXXX}}

\begin{document}

\maketitle
\thispagestyle{firststyle}
\ifthenelse{\boolean{shortarticle}}{\ifthenelse{\boolean{singlecolumn}}{\abscontentformatted}{\abscontent}}{}

\dropcap{G}lasses are ubiquitous, yet a first principles theory explaining their properties remains elusive.
A snapshot of the configuration of molecules in a glass appears to be the same as that of a dense liquid, yet glasses are solids not liquids.  
The recently proposed replica theory of glasses has overcome this lack of a structural signature and provides a first principles theory ~\cite{gardner_spin_1985, brito_geometric_2009,  parisi_mean-field_2010, charbonneau_fractal_2014, franz_universal_2015, biroli_breakdown_2016, charbonneau_glass_2017, berthier_gardner_2019}.  This theory predicts two distinct glass phases: a stable glass phase, characterized by an energy landscape of multiple disconnected local minima and a marginal glass phase characterized by an energy landscape of basins broken into sub-basins which are themselves broken into sub-basins, ad infinitum.
The marginal phase has been observed in multiple simulations \cite{charbonneau_numerical_2015, berthier_growing_2016, jin_exploring_2017, hicks_gardner_2018, seoane_spin-glass-like_2018, scalliet_marginally_2019}, however, it is not known whether the replica theory is applicable to an experimental glass.
Here we show that such a phase does exist in a slowly densifying colloidal glass, providing thermal experimental confirmation of the replica theory of glasses.
We use the mean squared displacement of individual colloids at extremely short time and length scales as an assay to determine the phase of matter.  
By observing as a function of density we find evidence for a reentrant marginal phase and can begin to map out the phase diagram for colloidal glasses and compare it to the mean field result~\cite{ biroli_breakdown_2016,scalliet_marginally_2019}.
Our results demonstrate that colloidal glasses self organize into a hierarchy of sub-basins characteristic of a marginal phase.  The existence of which provides evidence for a series of thermodynamic phase transitions underlying the dramatic change in behavior seen between a liquid and a glass.
This work provides a blueprint to creating a complete phase diagram for colloidal glasses, including behavior deep within the glassy phase.  The techniques developed in this work explore a previously underlooked regime for glasses, that of the shortest time and length scales. Measuring behavior in this regime is a potent tool with which to explore previously inaccessible implications of the mean field theories of glasses and to probe new material properties.

Within the replica theory of glasses, the energy landscape of a stable glass is dominated by distinct local minima contained within smooth basins.  A signature for this phase can be found in the mean squared displacement (MSD) of individual particles.   As in all thermal systems, at shortest times particles will move ballistically.  Because the basin is smooth, particles will rapidly explore their local cage and the MSD will reach a flat plateau characterized by the size of the cage.   This signature of the stable glass has been verified both computationally and experimentally \cite{pusey_observation_1987, mason_linear_1995, van_megen_measurement_1998, weeks_three-dimensional_2000,kegel_direct_2000,weeks_subdiffusion_2002, kaufman_direct_2006, hunter_physics_2012, kim_diffusing_2019}.

\begin{figure}
\includegraphics[scale=0.95]{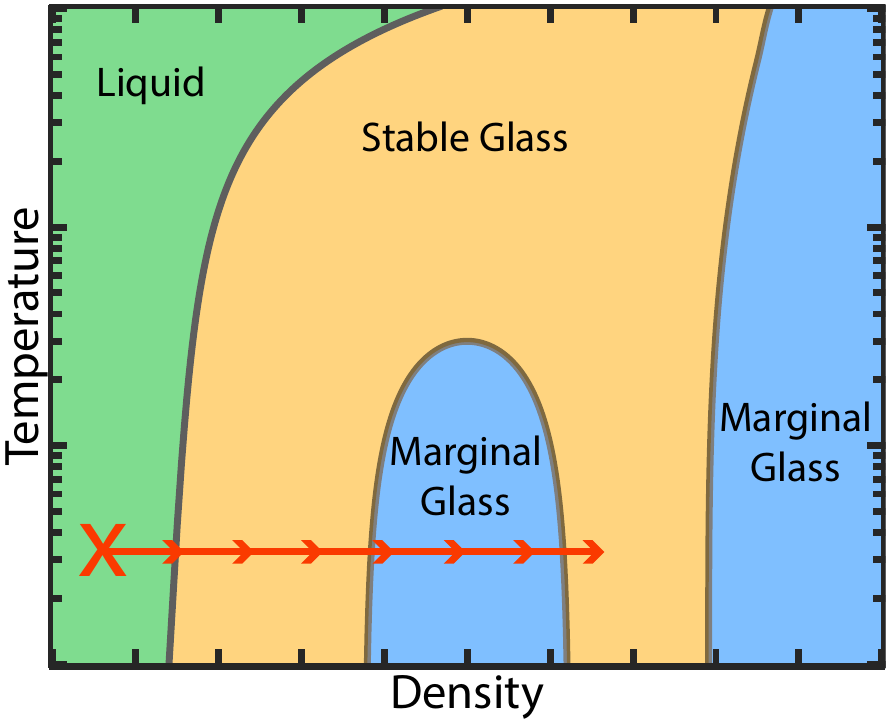}
\caption{Sketch of a phase diagram for soft sphere glasses as a function of temperature and density, adapted from ~\cite{ biroli_breakdown_2016,scalliet_marginally_2019}. Our experiment starts at the "X" and follows the arrows from the liquid to stable glass to marginal glass and then further into the stable glass phase.}
\label{PhsD}
\end{figure}

The energy landscape of a glass in the marginal (also known as a Gardner) phase is superficially similar to that of a stable glass in that it is non-ergodic and thus broken up into multiple independent basins.  However, the energy basin in a marginal glass is itself further broken up into a "fractally dense" hierarchy of sub-basins within sub-basins~\cite{hentschel_athermal_2011,franz_universal_2015,procaccia_breakdown_2016,biroli_breakdown_2016,jin_exploring_2017}.  Resultingly, a stable glass at any instant in time will necessarily be found in a single particular sub-basin with vanishingly small barriers to a set of nearby sub-basins contained within a higher level of sub-basin.  Thus, as the system evolves it will first efficiently explore the local sub-basin until such time as it experiences a large enough fluctuation to transition into a higher order sub-basin.  For longer times it will explore this higher order sub-basin, contained within a yet larger energy barrier.  

As time increases the system will explore progressively larger regions of the energy landscape that are separated from the rest by progressively larger energy barriers.  Each time the system breaks out of a sub-basin into a higher order sub-basin the MSD will be characterized by growth, as the system suddenly has access to more of phase space, followed by a plateau as the system fully explores this newly accessible space.  Due to the heirarchical nature of the energy landscape the system will exhibit ever longer plateaus at ever greater lengthscales.  In the thermodynamic limit this series of discrete plateaus will merge into a continuous logarithmic growth of the MSD~\cite{charbonneau_glass_2017, berthier_gardner_2019}. 

For soft spheres (i.e. particles with non-infinite resistance to deformation), the glass phase diagram (sketched in figure \ref{PhsD}) has been proposed to exhibit a re-entrant stable glass phase~\cite{ biroli_breakdown_2016,scalliet_marginally_2019}.  The athermal jamming density controls the location of  the onset of marginality in phase space and thus controls the density at which the marginal phase is first encountered, forming a so-called Gardner dome about the athermal jamming point~\cite{ biroli_breakdown_2016,rainone_following_2016,scalliet_marginally_2019}. That the system is able to explore densities above jamming is due to the soft sphere interaction and the presence of thermal excitatory energy.

The existence of the marginal glass regime and its transition out of the stable glass regime has been extensively probed with numerical simulations.  Examining hard and soft sphere systems, several of these simulations confirm the existence of the marginal glass~\cite{jin_exploring_2017} with some even including numerical MSDs with logarithmic-like scalings~\cite{charbonneau_numerical_2015,berthier_growing_2016, seoane_spin-glass-like_2018}.  However, other work using different algorithmic approaches contest this assesment, positing that the marginal phase is an avoided transition in low spatial dimensions~\cite{scalliet_absence_2017,hicks_gardner_2018}. Further, due to the high computational cost in numerical simulations, the transition from ballistic to logarithmic behavior in the MSD has never been observed. 

Probing such a marginal phase in physical glasses is difficult, necessitating novel methods.  The first experimental evidence compatible with the marginal glass phase was work done on an externally controlled pseudo-thermal two-dimensional disk system~\cite{ seguin_experimental_2016}.  That work was done by examining a direct replica signature similar to that employed in numerical simulations.  Evidence for the Gardner phase was found by resetting the system to identical starting points and observing random quasi-thermal excitations to evolve the system's energy landscape into a hierarchy of sub-basins.   A second work, studying the Johari-Goldstein Relaxation in sorbitol and xylitol under cooling to low temperature found an extreme broadening of the $\beta$ relaxation distribution, consistent with the fractal roughening of the energy landscape predicted by the marginal phase~\cite{geirhos_johari-goldstein_2018}. 


In this work we create a three dimensional thermal colloidal glass and follow its behavior under densification.  We use the MSD as a signature to characterize the glass phase as a function of elapsed time.  We observe a clear signature of both the stable and marginal phases.  We then qualitatively compare these results to the re-entrant glassy phase diagram for colloids~\cite{ biroli_breakdown_2016,scalliet_marginally_2019} (Figure \ref{PhsD}) and find good agreement.

\section{Methods}

\begin{figure}
\includegraphics[scale=1.0]{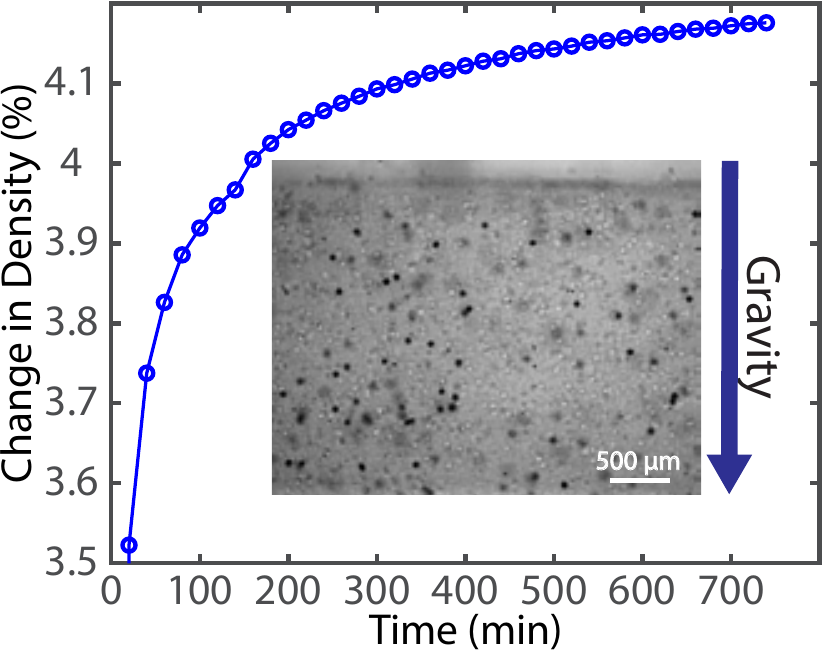}
\caption{Global change in density relative to the initially shaken sample versus time for a representative colloidal system.  Even at the longest times studied the system is still slowly densifying. Inset: Image of the colloidal glass during sedimentation.  The interstitial fluid has been index matched to the PMMA beads so they are nearly invisible. The black tracer particles are thus revealed.}
\label{methodsFig}
\end{figure}

	Our experimental system consists of nominally 50 $\mu$m diameter PMMA colloids (Cospheric PMPMS-1.2 45-53um $>95\%$) with 95\% falling within a range of 45-53 $\mu$m,  sphericity greater than 99\%, and density 1.20 g/cm$^3$.  The degree of polydispersity is chosen to frustrate crystallization.  These colloids are suspended in a mixture of tetrahydronapthalene, decahydronapthalene, and cyclohexyl bromide.  This three component liquid is chosen to allow for precise independent control of both the refractive index and density.  To this are added a small number of dyed black polyethylene tracer colloids with the same density and diameter as the PMMA colloids (Cospheric BKPMS-1.2 45-53 um). This colloidal suspension is placed into a sealed cuvette sample chamber with an imaging width of 3 mm, large enough to allow for imaging far from any boundaries.  The sealed cuvette allows us to reset the system to a new configuration by shaking it.
	
\begin{figure}[t]
\includegraphics[scale=1.6]{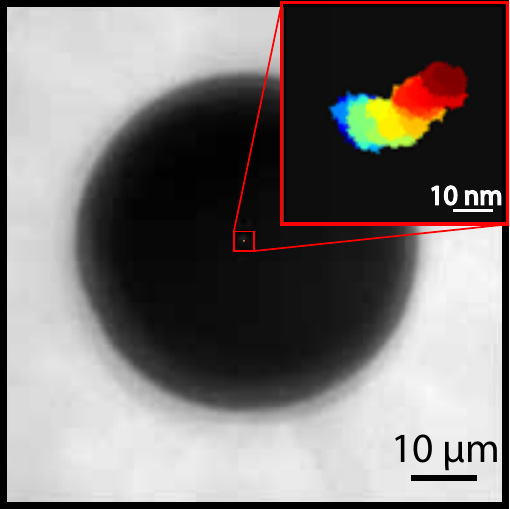}
\caption{Representative still image from the captured video showing the $\sim 50 \mu \text{m}$ diameter tracer-colloid inside the index matched colloidal glass. The tracked motion of the colloid for the entire 1.72 s video is overlayed on top of the colloid and blown up within the red box. The progression of time is represented by color running from start (blue) to end (dark red).}
\label{traceFig}
\end{figure}

	All imaging data was collected on a Nikon TE2000s microscope on a floating stage optical table in a climate controlled room.  Illumination is provided by a high intensity red LED (Thorlabs M660D2) with output of 940 mW.  High speed video was recorded on a Phantom Miro M310 high speed camera running at 64,000 fps using an image size of 192x192 pixels with a total collection of 110000 frames and a time of 1.72 s.  We used a 50x lens (Nikon LU plan ELWD 50x/0.55 B $\infty$/0 WD 10.1) giving us an image resolution of $0.4 \mu  \text{m}$ per pixel. The cuvette being imaged is stabilized on the microscope using a custom sample holder that ensured a flat imaging plane and a minimum of vibration.  The imaging stage is then encased in a small cardboard box for thermal and acoustic isolation.

	We match the solutions' index to the PMMA, but introduce a slight density mismatch betweens the fluid and the colloids causing the colloids to slowly sediment, resulting in a colloidal glass whose packing fraction increases with increasing time.  We measure the average sedimentation across many samples by imaging the whole cuvette as a function of time, as shown in figure \ref{methodsFig} (Inset).  We use particle image velocimetry to analyze the flow \cite{thielicke_PIVlab_2014} to compute the change in density versus time, a representative curve for which is shown in figure \ref{methodsFig}. The colloidal density always increases monotonically from the initial shaken state.  The rate of densification slows over time but never stops. After the first hour the colloids have mainly fallen out of the solution and begin to compact more slowly. The total change in packing fraction over the course of 700 minutes of sedimentation is about 4$\%$.

	Due to the extreme vibration sensitivity of these measurements all data is collected late at night to ensure a minimum of human induced noise.   Before imaging, the sample cuvette is vigorously shaken from multiple different directions in order to homogenize the colloidal suspension.  Immediately afterwards it is placed in the cuvette holder on the microscope at which point a suitable tracer-colloid is selected and centered in the field of view of the camera.  Tracer colloids are chosen no closer than 250 $\mu \text{m}$ from the cuvette inner walls (approximately the diameter of 5 colloids away). Automated high speed video of the colloid is recorded every twenty minutes.  Over the course of the first hour the tracer-colloid moves sufficiently that it is necessary to periodically move the stage to recenter the colloid. This motion is consistent with the large change seen in the sedimentation analysis.  However, after reaching a sufficiently dense system, the tracer-colloid moves so little that the stage is locked down for all subsequent videos.

\begin{figure}[tb!]
\includegraphics[scale=0.65]{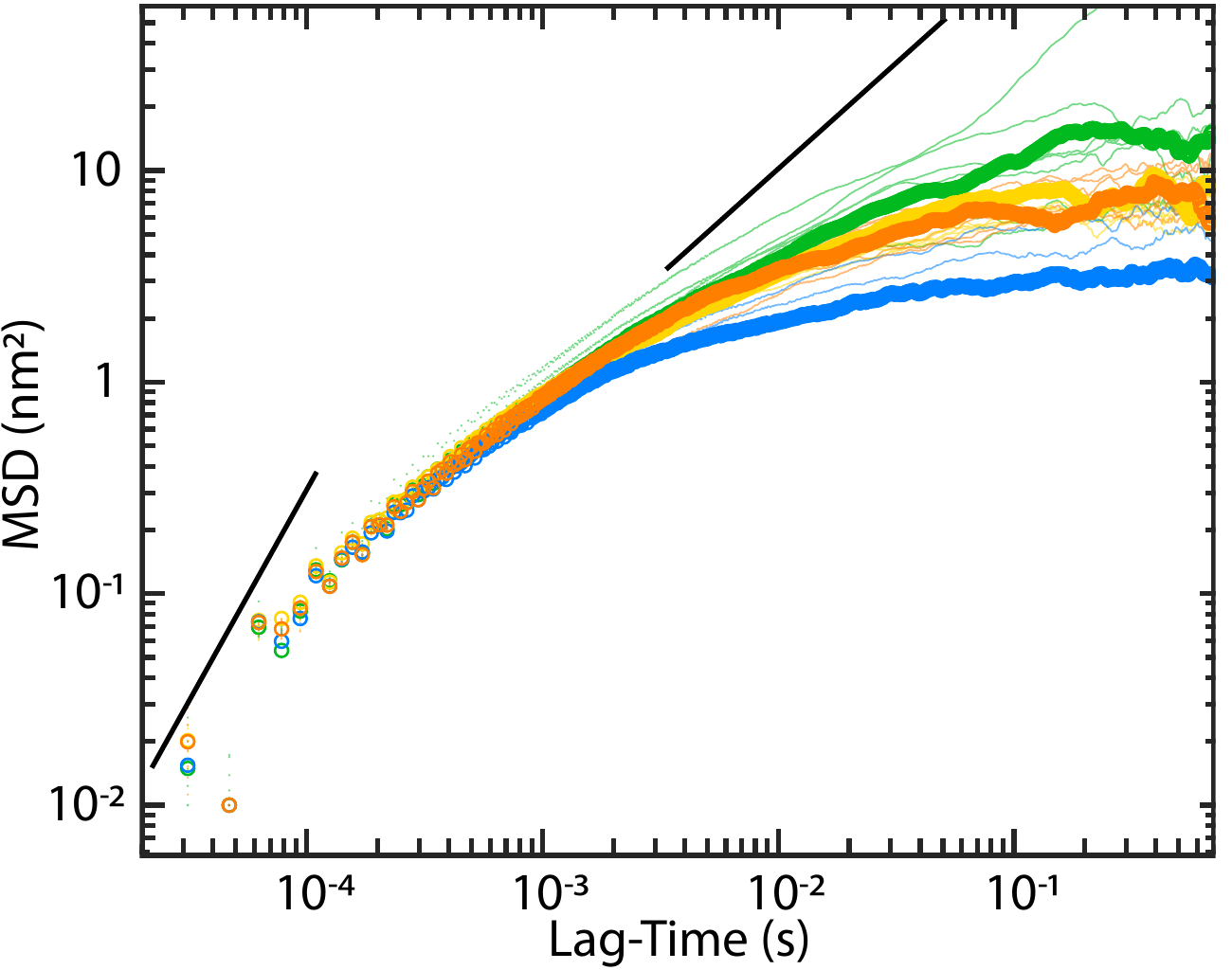}
\caption{MSDs from a sedimenting tracer-colloid plotted on a log-log scale.  The MSDs were taken every 20 minutes.  As a guide to the eye, we show ballistic and diffusive asymptotes demonstrating that at short times the motion is near ballistic and at long times the curves are all sub-diffusive.  There are four important MSD regimes with a representative highlighted: from shaken to 160 minutes the MSD is sub-diffusive (green, 100 min), from 180 to 320 minutes the MSD is plateau (yellow, 200 min), between 320 and 400 minutes the MSD is logarithmic (lblue 400 min), from 420 to 520 minutes the MSD is plateau, which is nearly the same as the previous plateau (orange, 580 min).  
}
\label{VarsLogLog}
\end{figure}

	We use a radial center tracking algorithm \cite{parthasarathy_rapid_2012} to find the center of the colloids motion in the plane perpendicular to gravity for every frame of the video.  This tracking algorithm is unbiased and has a mean localization error of $\sim 1.5$ nm for each frame.  From this tracked data we compute the drift subtracted MSD for a given video as: 
\begin{equation}
\text{MSD}(\tau) = \langle \| \vec x(t+\tau)-\vec x(t) \| ^2 \rangle)-\| \langle \vec x(t+\tau)- \vec x(t) \rangle ^2 \|
\end{equation}
where $\vec x(t)$ is the measured position of the particle at time $t$, $\tau$ is the lag-time between position measurements, and angle brackets denote a time average.  Typically, the drift is an extremely small part of the MSD effecting only the long time (i.e. $\tau >0.1$ s) motion.  Because the localization error is independent and identically distributed we can subtract it from this MSD to achieve measurements below the nominal noise floor.  This technique was employed and verified in previous work on the crossover from ballistic to diffusive motion in a freely floating colloid ~\cite{hammond_direct_2017}.

In order to distinguish between different forms of the MSD we begin by finding the lag time $\tau^*$ at which the presence of other particles begins to impinge upon the motion of the tracer.  We characterize this as the time when each MSD deviates by more than 10\% from the Clerx-Schram expression for free floating colloidal motion \cite{clercx_brownian_1992}.  If the motion is logarithmic then beyond this lag time it should be well fit by
\begin{equation}
\text{MSD}(\tau) = a\times\log\left( \frac{\tau}{\tau^*} \right)+x^*
\end{equation} 
where $a$ characterizes the slope of the logarithm and $x^*$ is the value of the MSD at $\tau^*$.  

Data Availability:  All data discussed in the paper will be made available to readers upon request.

\section{Results and Discussion}

\begin{figure*}[th!]
\includegraphics[scale=1.0]{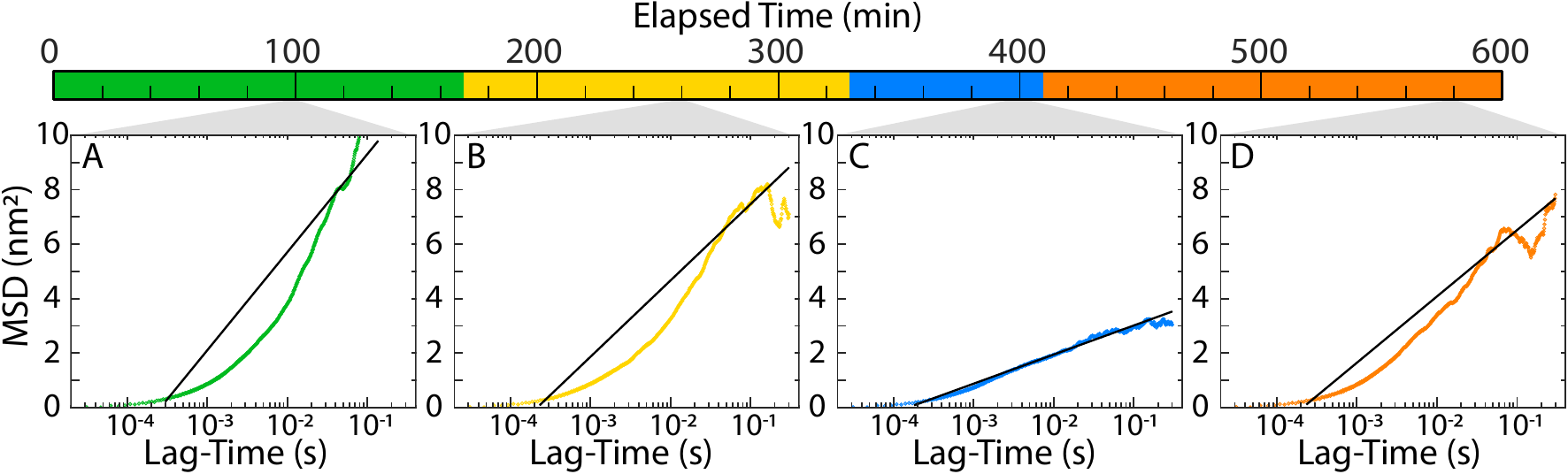}
\caption{Representative MSD curves from the four time steps. Superimposed on each MSD is the best fit logarithm. A: MSD in the sub-diffusive regime, taken at 100 minutes with a logarithmic fit. B: MSD in plateau regime, 260 minutes. C: MSD in logarithmic regime, 400 minutes. D: MSD in reentrant plateau regime, 580 minutes.}
\label{logfit}
\end{figure*}

	As shown in figure \ref{VarsLogLog}, we find a substantial change of behavior with increasing time and thus increasing density.  The  MSD of the tracer-colloid just after it has been shaken (green curves) shows ballistic motion at short times turning over to a subdiffusive motion at long times, consistent with that of a colloid in a dense liquid suspension.  At lag times of around $2\times 10^{-4}$s the MSD deviates from the expectation of a free floating thermal colloid~\cite{clercx_brownian_1992} and instead shows sub-diffusive motion~\cite{weeks_three-dimensional_2000}.  The absence of a diffusive regime results from the high initial colloidal density which leads to some caging even in a freshly shaken sample.  As seen in figure \ref{logfit}A these MSDs are clearly not logarithmic.

	As the colloidal density increases the power law of the sub-diffusive motion decreases towards zero.   Starting at about 180 minutes the MSD flattens to a plateau, indicating that the tracer-colloid is trapped inside a cage of other colloids. This behavior is shown in yellow in figure \ref{VarsLogLog} on a log-log scale and on a semi-log scale in figure \ref{logfit}B. This demonstrates that the system has entered into the stable glass phase.   The height of the plateaus show a small but overall consistent trend towards smaller cages, driven by the slow overall densification of the system.    The transition between ballistic motion and the plateau happens slowly, over nearly three decades, only reaching the flat plateau at $\tau \sim 1 \times 10^{-1}$s. Across all samples measured, the transition from sub-diffusive to plateau happens after about 160 minutes.

	After about 340 minute the constant plateau is replaced by an even lower MSD approaching that earlier plateau value, shown in blue in figure \ref{VarsLogLog}.  This change in the MSD is indicative of a change in phase of the underlying system.  As seen in figure \ref{logfit}C this motion is well fit by a logarithm.  These are the only curves to demonstrate a good logarithmic fit.  This behavior is indicative of the system entering the marginal glass phase.  In some systems, the logarithmic MSD as a function of starting time exhibits a progression towards smaller values of the slope fitting parameter.  This implies that small changes in packing fraction result in dramatic slowing down of dynamics.  Across all samples measured, the transition from plateau to logarithmic happens after about 380 minutes.

	The system remains in the marginal phase for about 80 minutes (across all samples, an average of 165 minutes).  After which the MSDs consistently cease to be logarithmic and pop back up to a constant plateau, indicating a return to the stable glass regime, shown in orange curves in figures \ref{VarsLogLog} and \ref{logfit}D.  There is no further evolution of the MSD during the experimental observation. In addition to the sedimenting tracer particle MSD shown in figures \ref{VarsLogLog} and \ref{logfit}, we have included similar data for other experimental runs in the supplemental material.

	Taken as a whole, the behavior of the densifying colloidal glass is consistent with the schematic phase diagram shown in Figure \ref{PhsD}.  The system begins in the liquid phase, at the point marked "X".  The experiment proceeds at a constant temperature but increasing density along the red line, passing through the stable glass phase, the marginal phase, and ending once again in a high density stable glass as shown in figure \ref{logfit}. It is striking that not only do we see individual signatures of each phase but that they combine to begin to map out a phase diagram which agrees with the theoretical predictions.

\section{Conclusions}

We have created a colloidal glass at fixed temperature and tracked it through the liquid, stable glass, and marginal glass phases as a function of increasing density.  We track an individual colloid deep within the glass at short times and find the associated MSD.  This experiment lays the groundwork for a complete experimental determination of the glass phase diagram by repeating such measurements over as broad a range of temperatures as possible.  A full phase diagram will also allow for a careful exploration of the transition between phases to characterize what, if any, phase transitions are present in colloidal glasses.  Further, the techniques developed in this work explore a previously underlooked regime for glasses, that of the shortest time and length scales. Measuring behavior in this regime is a potent tool with which to explore previously inaccessible implications of the mean field theories of glasses and to probe new material properties.  The experimental observation of these distinct phases provides evidence for the validity of the replica theory of glasses in physical colloidal glasses and opens the door to an exploration of its validity in the wider world of glassy materials.

%
%

\acknow{We thank Dan Blair, Raghuveer Parthasarathy, and Camille Scalliet for helpful discussions, and the University of Oregon machine and electrical shop staff.  This work was supported by National Science Foundation (NSF) Career Award DMR-1255370, and the Simons Foundation No. 454939.}

\showacknow{} 

\bibliography{ExperimentalGlassTrans,GardnerGlassPred,methods,sedimentation}

\end{document}